%Paper: hep-th/9209129
%From: FERRETTI%UORHEP.bitnet@CUNYVM.CUNY.EDU
%Date: Wed, 30 Sep 1992 15:14 EST

%%%%%%%%%%%%%%%%%%%%%%%%%%%%%%%%%%%%%%%%%
%%%                                   %%%
%%% PLAIN TEX FILE, NO MACROS NEEDED  %%%
%%%                                   %%%
%%%%%%%%%%%%%%%%%%%%%%%%%%%%%%%%%%%%%%%%%
\baselineskip=18pt
\magnification=1200
\def\tr{\;{\rm tr}\;}

UR-1278

ER-40685-730

hep-th/9209129

\centerline{THREE DIMENSIONAL QUANTUM CHROMODYNAMICS\footnote\dag
       {This work was supported in part by DOE Grant
       DE-FG02-91ER40685.}}
\vskip.5in
\centerline{G. Ferretti, S.G. Rajeev\footnote\ddag
        {Talk presented by S.G. Rajeev at the XXVI International
        Conference on High Energy Physics, Dallas TX Aug. 1992}
        and Z. Yang}
\centerline{Department of Physics and Astronomy}
\centerline{University of Rochester}
\centerline{Rochester NY 14627}
\vskip1in

\centerline{ABSTRACT}
\noindent The subject of this talk was the
review of our study of three ($2+1$) dimensional Quantum Chromodynamics.
In our previous works, we showed the existence of a phase
where parity is unbroken and the flavor group $U(2n)$ is broken to
a subgroup $U(n)\times U(n)$. We derived
the low energy effective action for the theory and showed that it has
solitonic excitations with Fermi statistic, to be identified with the three
dimensional ``baryon''. Finally, we studied the current algebra for this
effective action and we found a co-homologically non trivial generalization
of Kac-Moody algebras to three dimensions.
\vfill\eject

\centerline{INTRODUCTION}

Quantum Chromodynamics (QCD) is the universally accepted theory of strong
interactions. In spite of this fact, there are still many unresolved issues
that need to be addressed before we can declare our understanding of QCD
complete. Among the least understood problems are those that cannot
be studied by perturbation theory, such as chiral symmetry breaking
and quark confinement. It is therefore
natural to look for other models that retain the basic features
of QCD but allow one to study these issues in a simpler setting.
One way to construct such models is to lower the dimensionality of the
system.

Two ($1+1$) dimensional QCD has been extensively studied but it
fails to be a good analogue for some purposes:
gauge symmetries in two dimensions are
somewhat trivial and no spontaneous
breaking of continuous symmetries can
occur. We chose to consider three dimensional QCD and to study the
breaking of the global symmetries in this context. We find a remarkable
similarity between this model and what is known about
four dimensional QCD from the study of the Skyrme model.
Working in lower dimensions we have the
extra bonus of obtaining an effective theory that is more tractable. This
theory can be regarded as a limiting case of coset models that are
renormalizable in the $1/N$ expansions.
Furthermore, its current algebra is a nontrivial abelian extension of the
naive algebra and it might give rise to interesting representation theory.

\centerline{THREE DIMENSIONAL QCD}

Let us begin by writing down the Lagrangian density for three dimensional
QCD. The gauge group
is $SU(N_c)$, the tensor $F_{\mu\nu}$ is the curvature associated
to the gauge field and $q_i$ represent the quark fields. The flavor index $i$
runs from $1$ to $N$ and we will always assume $N$ to be even ($N=2n$).
Color and spinor indices are suppressed but it should be kept in mind that,
as spinor, $q_i$ is a two component complex Grassmann field.

In this notation, the Lagrangian reads
$$
    L=-{1\over\alpha}\tr F_{\mu\nu}F^{\mu\nu} + \sum_i\bar q_i
    (\gamma\cdot\nabla+m_i)q_i \eqno(1)
$$
In the massless limit, this Lagrangian possess a global $Z_2\times U(2n)$
symmetry. Since the number of flavor is even, it is possible to make all
quarks massive without breaking parity explicitly. This is done by pairing
them into doublets $(q_i,q_{-i})$ of equal and opposite masses,
$m_i=-m_{-i}$, $i=1\cdots n$.

Our two main results regarding the spontaneous breaking of the global
symmetries are the following. First, parity is not spontaneously broken
if $N$ is even. No Chern-Simons term for the color gauge field can arise and
the theory should be in the confining phase. The proof of this statement is
an adaptation of the argument given by Vafa and Witten for four dimensional
QCD$^2$. It relies on the possibility
of defining a positive fermionic measure
for the path integral. For completeness, let us remark that
this argument fails when the number of flavors is
odd$^3$, indicating that the theory is in a parity violating phase and
therefore it is not a good analogue for QCD.

Second, in the massless limit, the flavor symmetry $U(2n)$ is
spontaneously broken to $U(n)\times U(n)$. By arguments similar to those
given by Coleman and Witten$^4$, one can show that if the symmetry is
broken, then it must break to this specific subgroup. One is then left with
the task of finding at least one correlation function of the theory that
breaks the $U(2n)$ symmetry in the limit of vanishing quark masses.
This function can be chosen to be the two
point correlation function of the flavor currents:
$$
   <J_\mu^i(k)J_\nu^k(-k)>={{iN_c}\over{4\pi}}\epsilon^{ik}
   \epsilon_{\mu\nu\rho}k^\rho + O(k^2). \eqno(2)
$$
In the massless limit, the leading term of
this correlation function is proportional to
$\epsilon^{ik}=\pm\delta_{ik}$, the plus sign being present for $i=k>0$,
the minus sign being present in the opposite case. The coefficient in~(2)
can be calculated exactly from some no-renormalization theorems$^5$.

\centerline{EFFECTIVE LAGRANGIAN}

Having understood the symmetry breaking pattern, we can now write down the
effective Lagrangian$^1$ for the light particles of the theory, i.e. the
Goldstone bosons ``pions'' associated to the broken generators.
These are described by a field $\Phi$ valued in the Grassmannian manifold
${\rm Gr}_{2n,n}=U(2n)/U(n)\times U(n)$.
Earlier, a different coset model with a non trivial topological term
had been proposed by Rabinovici et al.$^6$, where the target manifold was
not the complex Grassmannian but the projective quaternionic space.

By the standard properties of the
Grassmannian, $\Phi$ can be regarded as a $2n\times 2n$ traceless
hermitian matrix satisfying the condition $\Phi^2=1$. The odd dimensional
analogue of the Wess-Zumino-Witten term (WZW) can also be found by studying
the generators of $H^4({\rm Gr}_{2n,n})$.
As in the four dimensional case, the WZW term cannot be written as the
integral of a local density but one can regard space-time $M_3$
as the boundary of a four dimensional manifold with boundary $M_4$.

The form of the effective action is, in differential geometry notation,
$$
   S={{F_\pi}\over 2}\int_{M_3}d\Phi*d\Phi+
   {{N_c}\over{64\pi}}\int_{M_4}\tr \Phi(d\Phi)^4. \eqno(3)
$$
The constant $F_\pi$ has the dimension of a mass in natural units and can
be regarded as the ``pion decay constant''. The particular form for the WZW
term (second term in eq.~(3)), has been chosen so that the equation of
motion derived from it possess the same discrete symmetry as three
dimensional QCD. In particular, the equation of motion must violate both
internal ($(t,x,y)\to(t,-x,y)$) and external ($\Phi\to-\Phi$) parity but
it must preserve the combination of both. The coefficient $N_c/64\pi$ in
front of the WZW term is fixed by requiring $\exp(-S)$ to be independent on
the extension to $M_4$ and by comparison with~(2).

One can relax the assumption that led to eq.~(3) and include the vector
mesons as dynamical degrees of freedom into the Lagrangian. It turns out
that this is the correct way to study the solitonic excitations
(``baryons'') of the theory. The vector mesons provide the short range
repulsion needed to stabilize the soliton.

In our theory, a Chern-Simons term for these vector mesons arises from the
WZW term.
This is the Chern-Simons term for the unbroken subgroup
$U(n)\times U(n)$, not to be
confused with the (unexisting) color
Chern-Simons term. Each $U(n)$ factor in
the unbroken subgroup gives rise to its own Chern-Simons term. The relative
coefficients of the two terms must be chosen to be equal and opposite in
order to preserve parity as a symmetry of the effective theory.
By writing $\Phi=\chi\epsilon\chi^\dagger$, $\chi\in U(2n)$, $\epsilon$
as in~(2) and denoting by $A_1$ and $A_2$ the vector fields associated to
each $U(n)$, the new effective action becomes
$$
 \tilde S ={{F_\pi}\over 2}\int\tr(\nabla_\mu\chi^\dagger\nabla^\mu\chi
             +m_\pi^2\epsilon\chi\epsilon\chi^\dagger) dx^3
             +{k\over{4\pi}}\sum_{i=1}^2(-1)^i
            \int\tr(A_idA_i+{2\over 3} A_i^3) \eqno(4)
$$
The two models~(3) and~(4) coincide in the limit of infinite vector meson
mass.

\centerline{BARYONS AS SOLITONS}

The ``baryons '' of three dimensional QCD are described by the
static solitonic solutions of eq~(4) with winding number one$^7$.
(Recall that $\pi_2({\rm Gr}_{2n,n})=Z$, the baryon number $B$
is always identified with the winding
number.) The low lying baryons transform
under flavor symmetry just as predicted by the quark model. The only
important difference is in the size of the baryon. The size predicted by
the sigma model is larger by a factor $\log(F_\pi/N_c m_\pi)$ than the
prediction from the naive quark model. This seems to indicate that three is
the lowest critical dimension for the applicability of the Skyrme model.

There are also higher baryon number solutions. Particularly pleasing is
the existence of a cylindrically symmetric $B=2$ ``di-baryon'' solution.
The existence of these solutions, in particular the stability of the
di-baryon solution has been proven by numerical methods, since the equation
of motion that arises for the cylindrically symmetric ansatz are not
exactly solvable. By use of relaxation methods we have obtained
the radial profile for the baryon density in both the $B=1$ and the
$B=2$ case$^7$.

\centerline{CURRENT ALGEBRA}

The effective action~(3) admits an interesting current algebra.
The canonical formulation of the WZW model yields the Kac-Moody
algebra in two dimension but fails to give a Lie algebra in four. In our
intermediate case we find that the current algebra (more precisely the
current-field algebra) is still a Lie algebra$^8$.
Care must be taken in choosing the
correct expression for the current $J$ in a way that makes the Poisson
brackets linear in $\Phi$ and $J$.
If the two dimensional space-like surface $\Sigma$ is a torus,
we can  write these relations in a plane wave basis:
$$\eqalign{
   &\{\Phi^a_m,\Phi^b_n\}=0,\quad\{J^a_m,\Phi^b_n\}=f^{abc}\Phi^c_{m+n}\cr
   &\{J^a_m,J^b_n\}=f^{abc}J^c_{m+n}-
   {k\over 16\pi} d^{abc}\epsilon^{ij}m_in_j\Phi^c_{m+n}\cr} \eqno(5)
$$
Above, $m,n$ are two-dimensional vectors with integer components.
Also, $d^{abc}$ is the usual symmetric cubic invariant of $U(2n)$ and
$f^{abc}$ the structure constants.

The canonical formulation is completed by two first class constraints
$$
  \Phi^2-1=0\quad
  \left[J,\Phi\right]_+ +{k\over 16\pi}\epsilon^{ij}
  (\partial_i\Phi\partial_j\Phi)=0, \eqno(6)
$$
and by the Hamiltonian function
$$
 H={1\over 2}\int_{\Sigma}\tr\bigg(-{1\over{F_\pi\sqrt{g}}}
 \big(J+{k\over 32\pi}\epsilon^{ij}\partial_i\Phi
 \partial_j\Phi\big)^2
 +{{F_\pi\sqrt{g}}\over 4}g^{ij}\partial_i\Phi\partial_j\Phi\bigg)d^2x.
 \eqno(7)
$$

It should be stressed that the quantities $J$ and $\Phi$ are invariant
under the action of ``gauge transformations'' generated by the constraints.
The constraints themselves express the fact that the co-adjoint orbit of
this algebra is the cotangent bundle of the Grassmannian.

We could quantize the action~(3) if we could find a unitary, highest weight
representation for this algebra. At first sight this seems
rather unphysical because~(3) is not perturbatively renormalizable.
However,~(3) belongs to a class of models that are renormalizable in the
$1/N$ expansion. This is more clear if we consider a less symmetrical
Grassmannian, where $\Phi$ takes values in
${\rm Gr}_{N,n}U(N)/U(N-n)\times U(n)$. For $n=1$ this is just the usual
$CP^{N-1}$ model, known to be renormalizable in the large $N$ limit.
For $n>1$, this model is still renormalizable in the limit $N\to\infty$,
$n$ finite. Its current algebra is still given by (5) and everything is
left unchanged, except that now $\tr\Phi=N-2n\not=0$.
What fails in this case is the connection with three
dimensional QCD because for these ``asymmetrical'' models it is not
possible to preserve parity. We expect these models to yield a good quantum
theory and this would be very exciting by itself.
The original model on ${\rm Gr}_{2n,n}$ can
probably be approached as limiting case.
\vfill\eject

\centerline{REFERENCES}

\item{[1.]} G. Ferretti, S.G. Rajeev and Z. Yang, ``The Effective
Lagrangian of Three Dimensional Quantum Chromodynamics''
UR1255 {\it Int. J. Mod. Phys.} to appear, (1992).

\item{[2.]} C. Vafa and E. Witten, ``Parity Conservation in Quantum
Chromodynamics'' {\it Phys. Rev. Lett.} 53 535-536 (1984).

\item{[3.]} A.N. Redlich, ``Gauge Noninvariance and Parity
Nonconservation
\hfil\break
of Three-Dimensional Fermions'' {\it Phys. Rev. Lett.}
52 18-21 (1984).

\item{[4.]} S. Coleman and E. Witten, ``Chiral-Symmetry Breakdown
in Large-N Chromodynamics'' {\it Phys. Rev. Lett.} 45 100-102 (1980).

\item{[5.]} S. Coleman and B. Hill, ``No More Corrections to the
Topological Mass Term in QED$_3$'' {\it Phys. Lett.} B159 184-188 (1985).
\hfil\break
G. Ferretti and S.G. Rajeev, ``$CP^{N-1}$ Model with a Chern-Simons Term''
{\it Mod. Phys. Lett.} A23 2087-2094 (1992).

\item{[6.]} E. Rabinovici, A. Schwimmer and S. Yankielowicz,
``Quantization in the Presence of Wess-Zumino Terms'' {\it Nucl. Phys.}
B248 523-535 (1984).

\item{[7.]} G. Ferretti, S.G. Rajeev and Z. Yang, ``Baryons as Solitons
in Three Dimensional Quantum Chromodynamics''
UR1256 {\it Int. J. Mod. Phys.} to appear, (1992).

\item{[8.]} G. Ferretti and S.G. Rajeev, ``Current Algebra in
Three Dimensions'' {\it Phys. Rev. Lett.} to appear, (1992).

\bye